\documentclass[12pt]{iopart}
\usepackage{iopams}  
\usepackage{url}
\usepackage{cite}
\usepackage{pdflscape}
\usepackage{lineno}
\begin{document}
\title{Neural blind deconvolution for deblurring and supersampling PSMA PET}

\author{\textmd{Caleb Sample}\textsuperscript{\textmd{1,2}}, \textmd{Arman Rahmim}\textsuperscript{\textmd{1,3,4}}, \textmd{Carlos Uribe}\textsuperscript{\textmd{3,4,5}}, \textmd{Fran\c{c}ois Benard}\textsuperscript{\textmd{3,4,6}}, \textmd{Jonn Wu}\textsuperscript{\textmd{7,8}}, \textmd{Roberto Fedrigo}\textsuperscript{\textmd{4,9}}, \textmd{Haley Clark}\textsuperscript{\textmd{1,2,8}}}
\vspace{0.1cm}
\address{\textsuperscript{1}Department of Physics and Astronomy, Faculty of Science, University of British Columbia, Vancouver, BC, CA\\
\textsuperscript{2}Department of Medical Physics, BC Cancer, Surrey, BC, CA\\
\textsuperscript{3}Department of Radiology, Faculty of Medicine, University of British Columbia, Vancouver, BC, CA\\
\textsuperscript{4}Department of Integrative Oncology, BC Cancer Research Institute, Vancouver, CA\\
\textsuperscript{5}Department of Functional Imaging, BC Cancer, Vancouver, BC, CA\\
\textsuperscript{6}Department of Molecular Oncology, BC Cancer, Vancouver, BC, CA\\
\textsuperscript{7}Department of Radiation Oncology, BC Cancer, Vancouver, BC, CA\\
\textsuperscript{8}Department of Surgery, Faculty of Medicine, University of British Columbia, Vancouver, BC, CA\\
\textsuperscript{9}Faculty of Medicine, University of British Columbia, Vancouver, BC, CA\\
Corresponding Author and Email: Caleb Sample, \texttt{csample@phas.ubc.ca}\\
}

\vspace{0.1cm}

\begin{abstract}
\textit{Objective}: To simultaneously deblur and supersample prostate specific membrane antigen (PSMA) positron emission tomography (PET) images using neural blind deconvolution. \textit{Approach}: Blind deconvolution is a method of estimating the hypothetical "deblurred" image along with the blur kernel (related to the point spread function) simultaneously. Traditional \textit{maximum a posteriori} blind deconvolution methods require stringent assumptions and suffer from convergence to a trivial solution. A method of modelling the deblurred image and kernel with independent neural networks, called "neural blind deconvolution" had demonstrated success for deblurring 2D natural images in 2020. In this work, we adapt neural blind deconvolution for PVE correction of PSMA PET images with simultaneous supersampling. We compare this methodology with several interpolation methods, using blind image quality metrics, and test the model's ability to predict kernels by re-running the model after applying artificial "pseudokernels" to deblurred images. The methodology was tested on a retrospective set of 30 prostate patients as well as phantom images containing spherical lesions of various volumes. \textit{Main Results}: Neural blind deconvolution led to improvements in image quality over other interpolation methods in terms of blind image quality metrics, recovery coefficients, and visual assessment. Predicted kernels were similar between patients, and the model accurately predicted several artificially-applied pseudokernels. Localization of activity in phantom spheres was improved after deblurring, allowing small lesions to be more accurately defined. \textit{Significance}: The intrinsically low spatial resolution of PSMA PET leads to PVEs which negatively impact uptake quantification in small regions. The proposed method can be used to mitigate this issue, and can be straightforwardly adapted for other imaging modalities.

\end{abstract}

\maketitle

\section{Introduction}
Prostate specific membrane antigen (PSMA) positron emission tomography (PET) is an imaging modality typically used to detect prostate cancer \cite{Afshar_2015}. PSMA ligands accumulate not only in prostate tissue, but also in the salivary glands, lacrimal glands, liver, spleen, kidneys, and colon \cite{trover, israeli, wolf,Schwarzenboeck2017}. This leads to undesirable dose to these healthy regions \cite{trover, israeli, wolf}. However, this typically unwanted uptake can be potentially utilized to investigate intra-salivary gland parenchyma, which is an area of active research in the context of radiotherapy treatment planning \cite{clark_2018, vanluijk_2015, buettner_2012}. 

Quantifying uptake statistics in the relatively small size of salivary glands is difficult, due to partial volume effects (PVEs) associated with the intrinsically low resolution of PET imaging \cite{marquis_2023}, which is typically over 5 mm \cite{einstein_2019}. PVEs in PET images primarily correspond to spill-over between image voxels due to the positron range and point-spread function of the scanner\cite{kjell_2012}. Patient motion and Poisson-distributed noise due to low photon counts further exacerbates blur in PET images. While official SNMMI guidelines recommend using the maximum standard uptake value ($SUV_{max}$) in single voxels for uptake metrics \cite{snmmi_guide}, maximum uptake values are affected by "spill-in" from neighbouring voxels and would be expected to differ within small volumes if PVEs could be mitigated. 

There have been attempts to mitigate partial volume effects in PET images using analytical methods \cite{kjell_2012, prev_2, prev_3, prev_4, hansen_2023}, with most traditional approaches involving estimation of the point spread function. These methods are not applicable in the case of collaborative studies involving multiple scanners, or where the point spread function is unknown. 

The ill-posed problem of simultaneously estimating the theoretical "deblurred" image, $\vect{x}$, along with the spread-function or blur kernel, $\vect{k}$, from the original image $\vect{y}$ ($\vect{y} = \vect{x} \ast \vect{k}$), is referred to as blind deconvolution. Traditional \textit{maximum a posteriori} (MAP)-based methods for solving blind deconvolution require estimation of prior distributions for the kernel and deblurred image and specialized optimization techniques to avoid convergence towards a trivial solution. MAP-based methods are governed by the equation
\begin{equation}
(\vect{k}, \vect{x}) = \arg \max_{\vect{k}, \vect{x}}\textrm{Pr}(\vect{k}, \vect{x} | \vect{y}) = \arg \max_{\vect{k}, \vect{x}}\textrm{Pr}(\vect{y} | \vect{k} , \vect{x}) \textrm{Pr}(\vect{x})\textrm{Pr}(\vect{k})
\end{equation}

where $\textrm{Pr}(\vect{y} | \vect{k} | \vect{x})$ is the fidelity term likelihood and $\textrm{Pr}(\vect{x})$ and $\textrm{Pr}(\vect{k})$ are the priors of the deblurred image and blur kernel, respectively. Many prior models have been suggested \cite{chan1998total, krishnan2011blind, levin2009understanding, liu2014blind, ren2016image, zuo2016learning}, but are generally hand-crafted and insufficient for accurate modelling of $\vect{x}$ and $\vect{k}$ \cite{ren_2020}.

In 2020, Ren et al. \cite{ren_2020} developed "neural blind deconvolution" for estimating the deblurred image and blur kernel from 2D natural images using two neural networks which are optimized simultaneously, as opposed to traditional MAP-based methods which generally employ alternating optimization to avoid a trivial solution\cite{ren_2020}. This is a self-supervised deep learning method that does not involve a separate training set, but instead learns to predict deblurred images independently for each image. Neural blind deconvolution out-performed \cite{ren_2020} other traditional methods based on prediction's peak signal-to-noise ratio, SSIM, and error ratio. 

In this work, we build off of the development of 2D neural blind deconvolution \cite{ren_2020, kotera_2021}, implementing changes to the network architecture and optimization procedures for suitability with 3D PSMA PET medical images. We also modify the architecture to accomodate simultaneous supersampling to enhance spatial resolution.

\section{Methods}

    \subsection{Retrospective Patient Data Set}
         Full-body [18F]DCFPyL PSMA PET/CT images were de-identified for 30 previous prostate cancer patients (Mean Age 68, Age Range 45-81; mean weight: 90kg, weight range 52kg-128kg). Patients had been scanned, two hours following intravenous injection, from the thighs to the top of the skull on a GE Discovery MI (DMI) scanner. The mean and standard deviation of the injected dose was $310 \pm 66$ MBq (minimum: 182 MBq, maximum: 442 MBq). PET images were reconstructed using VPFXS (OSEM with time-of-flight and point spread function corrections) (number of iterations/subsets unknown, pixel spacing: 2.73/3.16mm, slice thickness: 2.8/3.02mm). The scan duration was 180 s per bed position. For appropriate comparison of voxel values between patients, patient images were all resampled to a slice thickness of 2.8 and pixel spacing of 2.73 using linear interpolation. Helical CT scans were acquired on the same scanner (kVP: 120, pixel spacing: 0.98mm, slice thickness: 3.75mm). 
        
        Images were cropped to within 6 slices below the bottom of the submandibular glands and 6 slices above the top of the parotid glands. Cropping was employed to avoid exceeding time constraints and the 6 GB memory of the NVIDIA GeForce GTX 1060 GPU used for training. Registered CT images were used for delineating parotid and submandibular glands for calculating uptake statistics. Limbus AI \cite{limbus} was used for preliminary auto-segmentation of the glands on CT images, which were then manually refined by a single senior Radiation Oncologist, Jonn Wu. Contours were used to calculate uptake statistics within salivary glands. Images were normalized to the maximum voxel value prior to deblurring, and afterwards rescaled to standard uptake values normalized by lean body mass ($SUV_{lbm}$). Lean body mass was estimated from patient weight and height using the Hume formula \cite{hume1966prediction}.

    \subsection{Overview of Neural Blind Deconvolution}
        Neural blind deconvolution, as first demonstrated by Ren et al. \cite{ren_2020}, estimates a deblurred version, $\vect{x}$, of an actual image, $\vect{y}$, by training neural networks on a case-specific basis. Specifically, two neural networks, $G_x$ and $G_k$ are trained to predict $\vect{x}$, and a blur kernel, $\vect{k}$, whose convolution together yields a close estimate of the original image, $\vect{y}$. The use of neural networks for image prediction was motivated by the work of Ulyanov et al. on deep image priors \cite{Ulyanov_2020} which defines
        \begin{equation}
            G_x(\theta_x) = \vect{x} \,\,\,\, ,\,\,\,\, G_k(\theta_k) = \vect{k} \,\,\,\, , \,\,\,\, \vect{y} = \vect{x} \ast \vect{k}
        \end{equation}
        where $\theta_x$ and $\theta_k$ represent trainable model parameters of $G_x$ and $G_k$, respectively. The following implicit constraints exist on network outputs which are satisfied automatically due to the networks' architectures.
        \begin{equation}
            0 \le G_x(\theta_x) \le 1 \,\,,\,\, G_k(\theta_k) \ge 0 \,\,,\,\, \sum_i G_k(\theta_k)_i = 1
        \end{equation}
        Previously, \cite{ren_2020} \cite{kotera_2021} such networks were trained using a loss function including a mean squared error (MSE) fidelity term along with regularization terms, $R(\vect{x}, \vect{k})$, where
        \begin{equation}
            \vect{x}, \vect{k} = \arg \min_{\vect{x}, \vect{k}} || \vect{x} \ast \vect{k} - \vect{y}||_2^2 + R(\vect{x}, \vect{k})
        \end{equation}

        The mathematical formulation of our problem differs slightly as we have adapted $G_x$ to yield an output image, $\vect{x}$ which is twice as large as $\vect{y}$ in each dimension to accomplish supersampling. Our $\vect{x}$ is then linearly downsampled to the original image size, $\vect{x_{\downarrow}}$, before being convolved with the kernel. Furthermore, we incorporate a component of mean absolute error into our loss function's fidelity term.
        
        \begin{equation}   
                \vect{x}, \vect{k} = \arg \min_{\vect{x}, \vect{k}} || \vect{x_{\downarrow}} \ast \vect{k} - \vect{y}||_2^2 +  || \vect{x_{\downarrow}} \ast \vect{k} - \vect{y}||_1 +  R(\vect{x}, \vect{k})  
        \end{equation}
        This can be written in terms of the model parameters $\theta_x$ and $\theta_k$ as 
        \begin{equation}   
                \theta_x, \theta_k = \arg \min_{\theta_x, \theta_k} || G_x(\theta_x)_{\downarrow} \ast G_k(\theta_k) - \vect{y}||_2^2 +  || G_x(\theta_x)_{\downarrow} \ast G_k(\theta_k) - \vect{y}||_1 + R( G_x(\theta_x), G_k(\theta_k) )  
        \end{equation}
        which can then be optimized via backpropagation of loss gradients.
    \subsection{Model Architecture}   
\begin{figure}[h]
      \centering
      \includegraphics[width=1.1\textwidth]{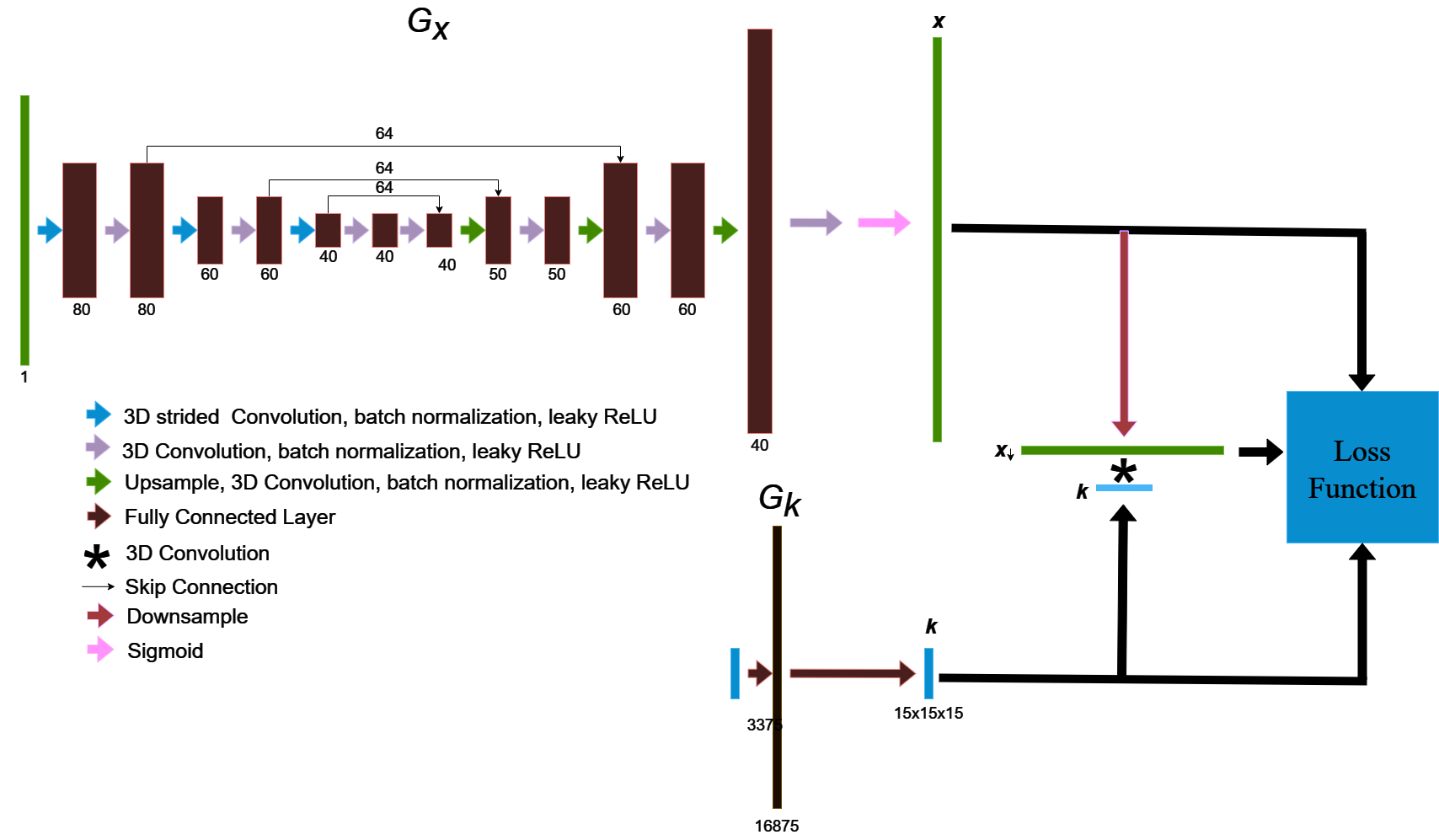}
      \caption{The blind deconvolution architecture used for deblurring PSMA PET images is illustrated. $G_x$ is an asymmetric, convolutional auto-encoder network for predicting the deblurred image, $\vect{x}$, and $G_k$ is a fully-connected network for predicting the blur kernel, $\vect{k}$. The convolution of $\vect{x}$ and $\vect{k}$ is trained to match the original image by using back-propagated model gradients from the loss function for iterative optimization.}
      \label{fig:architecture}
    \end{figure}

        Neural blind deconvolution was originally demonstrated on 2D images using a "U-Net"\cite{ronneberger2015unet}-style, symmetric, multiscale, autoencoder network for $G_x$ and a shallow fully-connected network for $G_k$. We built off of this methodology while implementing some notable changes to optimize model performance on PSMA PET images. A schematic diagram of the network design is shown in Figure~\ref{fig:architecture}.

        \subsubsection{$\mathbf{G_x}$ Architecture}
        
        All operations involving $G_x$ were adapted for use with 3D images. The network includes only 3 layers in the encoder-decoder series (as opposed to 5 in Ren et al.'s implementation) \cite{ren_2020}. A decrease in network layers is justified by the low resolution of PET images, which requires a smaller receptive field than needed for high-resolution images to localize edge features \cite{horwath_2019}. An additional layer was added to the end of the decoder section, including an upsampling operation, and double convolution, to output a final $\vect{x}$ twice as large in each dimension as the original input. This design allows $\vect{x}$ to be simultaneously supersampled and deblurred within the blind deconvolution process.

        Rather than iteratively increasing channels towards deeper regions of the encoder and decoder, we found that reversing the design such that the channel count decreases towards inward regions of the network produced favourable results. For each double convolution, the first convolution decreases the image size via a stride of 2. We included a high number of skip connection channels (64), as it was found to improve prediction accuracy.

        Furthermore, we scaled the final sigmoid layer used in previous versions of neural blind deconvolution \cite{ren_2020, kotera_2021}, as a regular sigmoid constrains the $\vect{x}$ to have the same maximum value as $\vect{y}$. This is not necessarily true, nor expected for the case of partial volume effect correction. We therefore scaled the final layer by $3/2$. 

        \subsubsection{$\mathbf{G_k}$ Architecture}    
            $G_k$ is a single layer perceptron, whose architecture Ren et al. \cite{ren_2020} found to outperform the auto-encoder style design of $G_x$ for the case of predicting the kernel. Model input is a 1-dimensional vector of length 3375, which is fed through a fully connected layer with 16,875 nodes, to an output of length 3375 which is passed through a SoftMax operation before being reshaped into a $15 \times 15 \times 15$ 3D kernel image. The softmax function ensures that kernel voxels sum to unity, and has the form: 
            \begin{equation}
            \sigma (\vect{k}_i) = \frac{e^{z_i}}{\sum_{j=1}^{3375}e^{z_j}}
            \end{equation}

    \subsection{Loss Function and Optimization}   

         It was unnecessary to form distinct training and validation sets as needed in supervised learning algorithms. Neural blind deconvolution is a type of ``zero shot'' \cite{Shocher_2018_CVPR} self-supervised learning algorithm, which utilises deep learning without requiring prior model training. Instead, parameters of $G_x$ and $G_k$ are optimized separately for each individual patient to predict the deblurred images. The algorithm for updating network weights to predict deblurred PSMA PET images is summarized in Table~\ref{tab:psma_alg}
    
\begin{table}[h]
            \centering
            \captionsetup{justification=raggedright}
            \caption{The optimization algorithm for updating network weights to predict deblurred PSMA PET images. This algorithm builds off of Ren et al.'s proposed joint optimization algorithm \cite{ren_2020} and implements modifications suggested by Kotera et al. \cite{kotera_2021}.}
            \footnotesize
            \begin{tabular}{@{}l}
            \br
            \multicolumn{1}{c}{\textbf{\large{Network Optimization Algorithm}}}\\
            \br
            \multicolumn{1}{l}{\textbf{Input}: Original PSMA PET image, $\vect{y}$}\\
            \multicolumn{1}{l}{\textbf{Output}: Deblurred PSMA PET image, $\vect{x}$, and blur kernel, $\vect{k}$}\\
            \br
            \textbf{Pre-training}\\
            1. Downsample $y$ to 1/2 resolution\\
            2. Initialize input $\vect{z_x}$ from uniform distribution to match size of $\vect{y}$\\
            3. Initialize $\vect{z_k}$ as size $7\times7\times7$ Gaussian kernel with standard deviation of two voxels\\
            4. \textbf{for}\verb| i = 1 to 1000:|\\
            5.     \texttt{~~~~$\vect{x} = G^i_x(\vect{z_x})$}\\
            6.     \texttt{~~~~$\vect{k} = G^i_k(\vect{z_k})$}\\
            7.     \texttt{~~~~Compute loss and back-propagate gradients}\\
            8.     \texttt{~~~~Update $G^i_x$ and $G^i_k$ using the ADAM optimizer \cite{adam}}\\
            9.     $\vect{x} = G^{1000}_x(\vect{z_x})$, $\vect{k} = G^{1000}_k(\vect{z_k})$\\
            10. Upsample $\vect{k}$ by 11/7 and downsample $\vect{x}$ by a factor of $\sqrt{2}$ \\
            11. Downsample $\vect{y}$ by $\sqrt{2}$ to match new convolution size\\
            12. $\vect{z_x} = \vect{x}$ , $\vect{z_k} = \vect{k}$\\
            13. Repeat steps 4 through 9.\\
            14. Upsample $\vect{k}$ by 15/11 and downsample $\vect{x}$ by a factor of $\sqrt{2}$ \\
            15. $\vect{z_x} = \vect{x}$ , $\vect{z_k} = \vect{k}$\\
            \br
            \textbf{Main Training} \\
            16. \textbf{for}\verb| i = 1 to 5000:|\\
            17.     \texttt{~~~~$\vect{x} = G^i_x(\vect{z_x})$}\\
            18.     \texttt{~~~~$\vect{k} = G^i_k(\vect{z_k})$}\\
            19.     \texttt{~~~~Compute loss and back-propagate gradients}\\
            20.     \texttt{~~~~Update $G^i_x$ and $G^i_k$ using the ADAM optimizer \cite{adam}}\\
            21.     $\vect{x} = G^{5000}_x(\vect{z_x})$, $\vect{k} = G^{5000}_k(\vect{z_k})$\\
            \br

        \end{tabular}\\
        \label{tab:psma_alg}

    \end{table}
        The main optimization procedure consisted of 5000 iterations broken into 3 stages, stage 1: $0 \le$ step $< 300$, stage 2: $300 \le$ step $< 2000$, stage 3: $2000 \le$ step $< 3500$, stage 4: step $\ge 3500$. 5000 iterations was chosen as it had been used in previous neural blind deonvolution methods, and furthermore because we found this number of iterations to be sufficient for the training loss to level out. We implemented a multi-scale pre-training procedure similar to the method recommended by Kotera et al. \cite{kotera_2021}. This involves initially training $G_x$ and $G_k$ to predict $\vect{x}$ and $\vect{k}$ using $2\times$ downsampled PSMA PET images, which are then upsampled by $\sqrt{2}$ and used as input for a second round of pre-training. The output is then upsampled back to the original size to serve as inputs for the regular training stage. We modify the procedure further such that after every upsampling during pre-training, the models, $G_x$ and $G_k$ are trained to predict their own inputs for 100 iterations, which was sufficient to allow training in the next scale to begin by predicting upsampled versions of the predictions in the previous layer. The size of $\vect{k}$ varies with the size of $\vect{x}$, such that three kernel sizes are used, $7\times7\times7$, $10\times10\times10$, and the original size $15\times 15\times 15$. For the downsampled pre-training of $\vect{x}$ and $\vect{k}$, 1000 iterations were used in each stage. This number of iterations was chosen to allow the pre-training image to approximately converge to its final estimate. The initial input vector for $G_k$ was a Gaussian Kernel with a standard deviation of two voxels, and the initial input vector for $G_x$ was a pseudorandom-generated array, sampled from a uniform distribution.
        The loss function components for each stage are summarized in Table 1. The parameters of $G_x$ and $G_k$ were updated using the ADAM optimizer \cite{adam}.

        \begin{table}[h]
        \centering
            \caption{Loss Function components used for optimization of $G_x(\theta_x)$ and $G_k(\theta_k)$. To guide training during the first 300 optimization steps, the Structural Similarity Index Measure (SSIM) is minimized between the deblurred image $\vect{x_\downarrow}$ and the registered CT's texture map. Conditions are invoked on the kernel estimate until the final 1500 steps of optimization. TV loss is introduced after the first 2000 steps, to avoid convergence towards a trivial solution. An $\ell^2$ norm loss term is applied to kernel, but only to voxels with values exceeding 0.7. }
            \footnotesize
            \begin{tabular}{@{}llll}
            \br
            $0 \le$ step $< 300$&$300 \le$ step $< 2000$&$2000 \le$ step $< 3500$&step $\ge 3500$\\
            \mr
            MSE + MAE &MSE + MAE & MSE + MAE& SSIM \\
            CT Texture SSIM & Kernel MSE& Kernel MSE& TV \\
            Kernel MSE& & TV & \\
            
            \br
        \end{tabular}\\

    \end{table}
    
     It was previously reported by \cite{kotera_2021} that adopting a total variation (TV) loss term biases outputs towards a trivial solution during the initial stages of training, but improves image quality as predictions near the final, correct solution. We experienced similar findings, and 
     therefore implemented a TV loss term only after the first 2000 iterations. After the first 3500 iterations, the fidelity term of the loss function transitions from a combination of MSE and MAE to the structural similarity index metric, which was also found to boost performance during final stages \cite{kotera_2021}. 
    
    During the first 300 iterations of training, we employ registered computed tomography (CT) images to provide anatomical guidance. This is implemented by including a SSIM loss term involving $\vect{x}$ and a CT texture map of the grey level run length matrix (GLRLM) with a long run emphasis, computed with pyradiomics \cite{pyradiomics}. We opted to use the texture map as opposed to the regular CT image as CT texture maps have been demonstrated to have a closer correlation with PSMA PET uptake \cite{sample_2023_hetero} than standard images. CT guidance is used to steer outputs towards the correct solution during the the first 300 steps of training before being dropped, to prevent competition with the fidelity term as outputs move closer towards final estimates. A previous study using traditional MAP-based blind deconvolution of PET images demonstrated the utility of magnetic resonance images for guiding optimization \cite{guerit2016blind, Gillman2023, song_2019}.

    A cumbersome challenge in performing a blind deconvolution is avoiding a trivial solution, $\vect{x} \ast \vect{k} =\vect{x} \ast \delta = \vect{y}$ \cite{Perrone_2014, Chaudhuri2014, yu_2019_trivial}. It is therefore necessary to impose regularization on kernel predictions. Previously, \cite{ren_2020, kotera_2021} applied an $\ell ^2$ norm constraint to the kernel to avoid convergence to a delta function. We also applied an $\ell ^2$ norm term; however, we only penalized kernel values larger than 0.7, to avoid a trivial solution while not penalizing small- to mid-range values in the kernel.

    \subsection{Patient Analysis}
        
        Ground truth deblurred patient images were not available, and we therefore adopted a variety of strategies and metrics for evaluating the quality of deblurred patient images and the model's ability to predict accurate blur kernels. 
    
        Two quantitative blind image quality metrics were compared between $\vect{x}$ and $\vect{y}$. First, the Blind/Referenceless Image Spatial Quality Evaluator (BRISQUE) score \cite{brisque} was used, which varies between 0 and 100, indicating the lowest and highest possible image quality, respectively. BRISQUE takes an input image and computes a set of features using the distribution of intensity and relationships between neighbouring voxels. It then predicts its image quality from its predicted deviation from a natural, undistorted image. Second, images were compared using the Contrastive Language-Image Pre-training (CLIP) metric \cite{clip} which varies between 0 and 1. CLIP is a multi-modal (language and vision) neural network trained with millions of image/text pairs. The model can be used for predicting both image quality and image caption quality. The clip network predicts image quality by ranking the applied captions "good photo" and "bad photo". CLIP was compared with numerous other quality metrics and was found to perform second to only BRISQUE. 

        The theoretical blur kernel of PET images has a component resulting from the point spread function of the scanner, as well as other limitations and patient motion. Since all PET images were acquired on the same scanner, with the same image collection protocol, it is expected that predicted blur kernels should have high inter-patient similarity,  Kernels were compared between patients quantitatively by taking the inner product of normalized kernel images. To validate the ability of neural blind deconvolution to accurately recover blur kernels, rather than arbitrary or semi-random shapes imposed by the loss function/optimization, we generated four types of "pseudokernels" and convolved them with previously deblurred images, which were then fed back into the blind deconvolution model. Predicted kernels were then compared with pseudokernels visually and by taking inner products. Generated pseudokernel shapes included a regular Gaussian, and 3 Gaussians skewed in the x, y, and z direction. 
        
    \subsection{Phantom Analysis}  

    The deblurring methodology was further tested using PET images acquired using the Quantitative PET Prostate Phantom (Q3P) \cite{fedrigo_2024}. The phantom had been developed for a previous study \cite{fedrigo2022quantitative} and included a set of ``shell-less'' spheres (3-16 mm diameters) of \textsuperscript{22}NaCl infused epoxy resin. The positron energy via \textsuperscript{22}Na decay is similar to that from \textsuperscript{18}F (220.3 keV and 252 keV respectively) \cite{jodal2014positron, jodal2012positron}, making \textsuperscript{22}Na suitable for approximating the decay seen in [18F]DCFPyL PET. The phantom contains eight spheres with diameters of 16 mm, 14 mm, 10 mm, 8 mm, 7 mm, 6 mm, 5 mm, and 3 mm arranged evenly in a coaxial ring and 50 mm from the radial centre of the phantom. The spheres were prepared with a \textsuperscript{22}Na concentration of 57.6 kBq/mL to represent lesion concentrations from an analysis of patients imaged with [18F]DCFPyL PET \cite{fedrigo2022quantitative}. By the time of acquisition, activity concentrations had decayed to approximately 31 kBq/mL. To establish realistic background concentration, the phantom was filled with a background concentration of 2 kBq/mL. To account for Poisson noise, the phantom was imaged for five noise realizations (2.5 min scan time) using a 5-ring Discovery MI PET/CT scanner. Phantom images were reconstructed using VPFXS (8 subsets, 4 iterations, 6.4 mm filter) using the clinical scanner console. The phantom specifications and image acquisition process are described in further detail by Fedrigo et al. \cite{fedrigo_2024}

    Phantom images were deblurred using the same neural blind deconvolution methodology applied to patient images. However, registered CT images of phantoms did not exist and therefore texture features could not be included during early stages of optimization. As regions of high uptake within the phantom were all spherically symmetric, there was inherent ambiguity in the extent of blurring. This is because multiple combinations of blur kernels and ``deblurred'' images can be convolved to produce highly similar observed images. This ambiguity is broken when shapes of arbitrary differing shapes are included in images, as in the case of patient acquisitions. To mitigate this challenge, we modified the final layer of $G_x$ to produce maximum values of 1.1 $\times$ the maximum voxel in the original image, rather than 1.5 $\times$. This did not present physical challenges, as the maximum voxel value was approximately equal to the known activity concentration in the largest sphere.

    Phantom spheres were contoured via two methods. First, a single fixed threshold that minimized the total difference between contoured and actual sphere volumes was determined separately for original and deblurred images. Second, variable thresholds were determined for each sphere size individually. Thresholds were expressed as a fraction of the maximum voxel value within a given region, and values between 0.2 and 0.8 (0.01 step size) were tested. Regions were defined by all voxels connected to the maximum voxel within a given sphere and greater than the given threshold. Actual volumes are referred to as $V_0$. The 3 mm sphere was not included in the analysis as it was poorly defined on original and deblurred images. 
    
    Activity concentration uniformity was first compared in original and deblurred sphere images by plotting intensity profiles. Recovery coefficients (RCs) for the mean, median, and maximum activity concentration ($RC_{mean}$, $RC_{median}, RC_{max}$) were also determined. These were defined as
    \begin{equation}
        RC_{stat} = \frac{a_{h, stat}}{a_{h, true}}
    \end{equation}
    where $a_{h, true}$ is the ground truth activity concentration for each sphere determined in a previous study \cite{fedrigo2022quantitative}, and $stat =$ \{mean, median, max\}. The mean inner product between combinations of phantom kernels was calculated. The difference in total activity within deblurred and original images was assessed.

\section{Results} 
\subsection{Patient Analysis}
Deblurred images had higher BRISQUE and CLIP scores than original images upsampled using nearest-neighbours, linear, quadratic, and cubic interpolation. These results are summarized in Table~\ref{tab:biq}.
\begin{table}[h]
            \centering
            \captionsetup{justification=raggedright}
            \caption{Blind image quality metrics, CLIP and BRISQUE are compared for original ($\vect{y}$) and deblurred ($\vect{x}$) PSMA PET images. For appropriate comparison, $y$ was upsampled by 2 in each dimension to match the size of $\vect{x}$. In the first column, $\vect{y}$ is upsampled using nearest neighbours interpolation to remain visually identical. We also compared the quality of $\vect{y}$ when upsampled using linear, quadratic and cubic interpolation ($\vect{y_{\uparrow_1}}$,$\vect{y_{\uparrow_2}}$,$\vect{y_{\uparrow_3}}$).}
            \footnotesize
            \begin{tabular}{@{}cccccc}
            \br
            &Original&\multicolumn{3}{c}{Polynomial Interpolation}&Proposed Method\\
            &$\vect{y}$& $\vect{y_{\uparrow_1}}$& $\vect{y_{\uparrow_2}}$&$\vect{y_{\uparrow_3}}$&$\vect{x}$\\
            \mr
            CLIP&$0.19 \pm 0.03$&$0.33 \pm 0.05$&$0.32 \pm 0.05$&$0.32 \pm 0.05 $&$0.39 \pm 0.04$\\
            BRISQUE&$75.8 \pm 8.1 $&$89.6 \pm 5.4 $&$85.0 \pm 6.3$&$84.8 \pm 6.32$&$90.5 \pm 7.5$\\
            
            \br
        \end{tabular}\\
        \label{tab:biq}

    \end{table}
    
    \begin{figure}[h]
      \centering
      \includegraphics[width=1.1\textwidth]{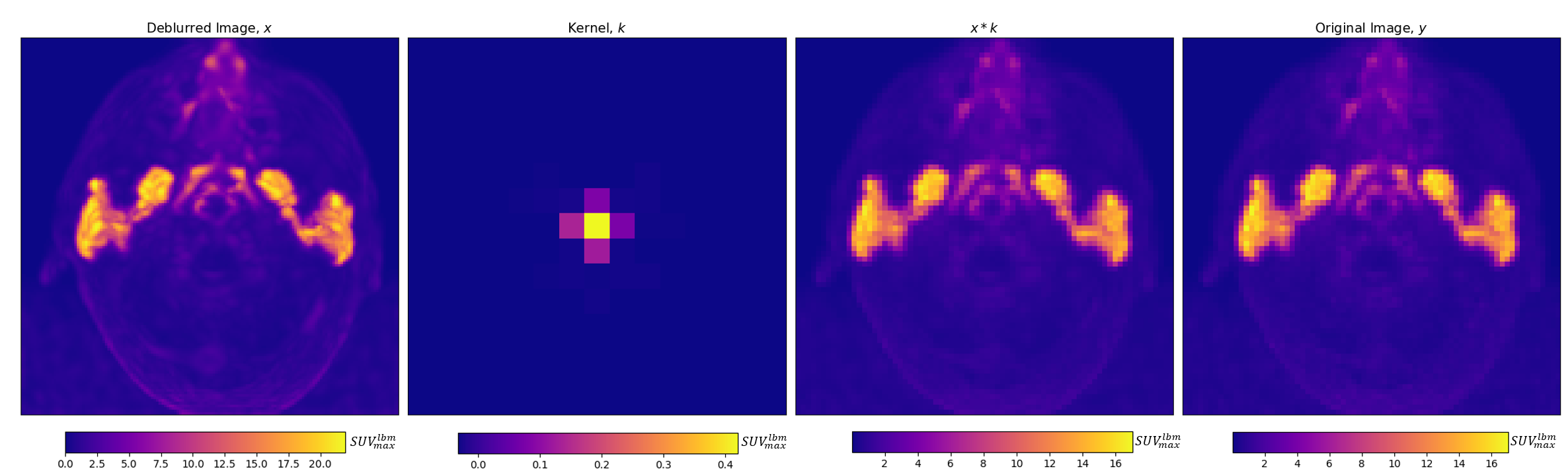}
      \caption{From left to right, for an individual patient, a deblurred maximum intensity projection image, and axial projection of the predicted kernel are shown, along with the maximum intensity projection of the convolution of the deblurred image and kernel, which is trained to match the original image, on the right. The inferior aspect of the tubarial glands are visible between the parotid glands.}
      \label{fig:x_k_conv_y}
    \end{figure}

\begin{figure}[h]
      \centering
      \includegraphics[width=1.1\textwidth]{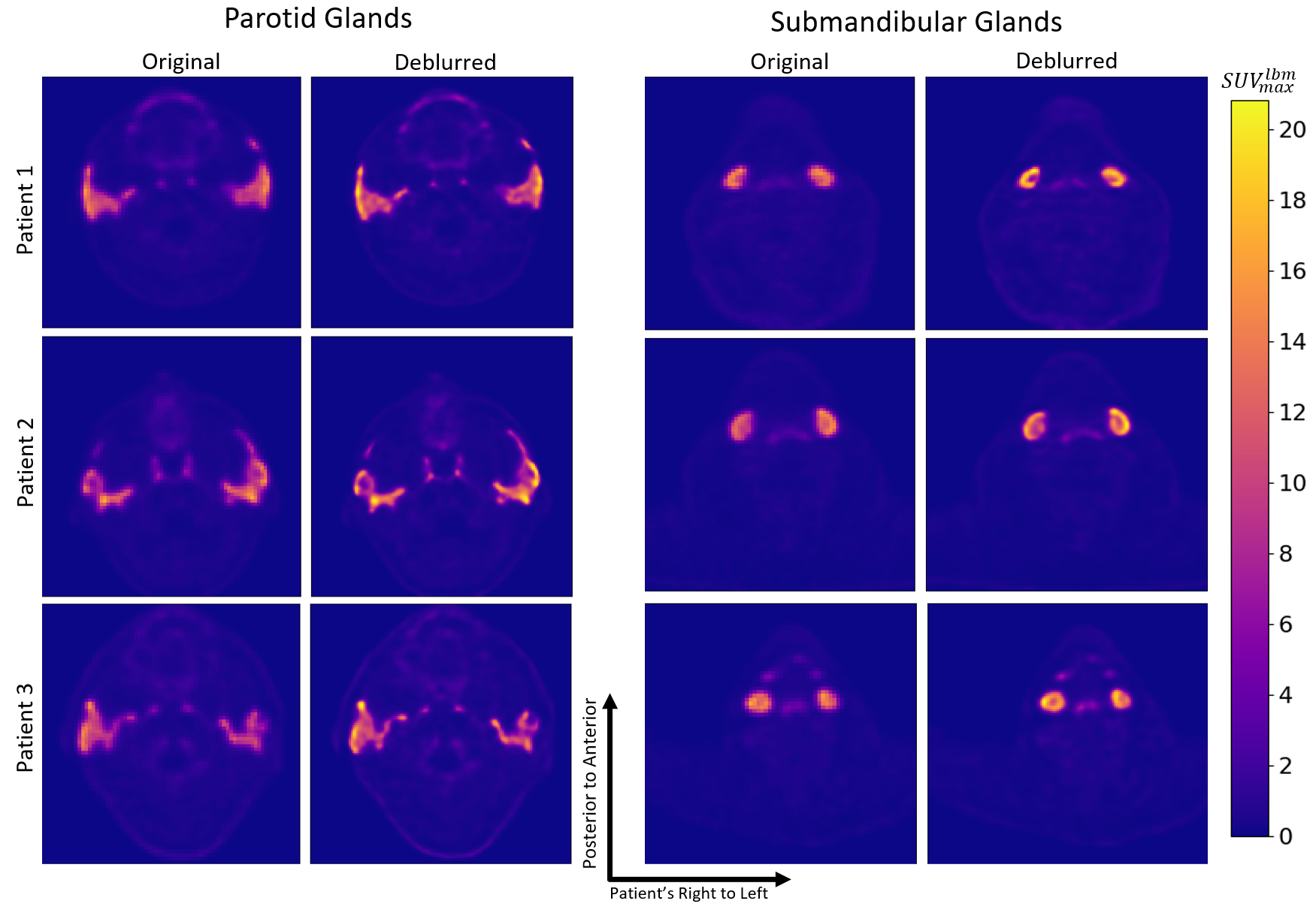}
      \caption{Axial slices intersecting parotid glands (left) and submandibular glands (right) for 3 different patients are shown on original and deblurred images.}
      \label{fig:par_sm_comp_axial}
    \end{figure}

\begin{figure}[h]
      \centering
      \includegraphics[width=0.8\textwidth]{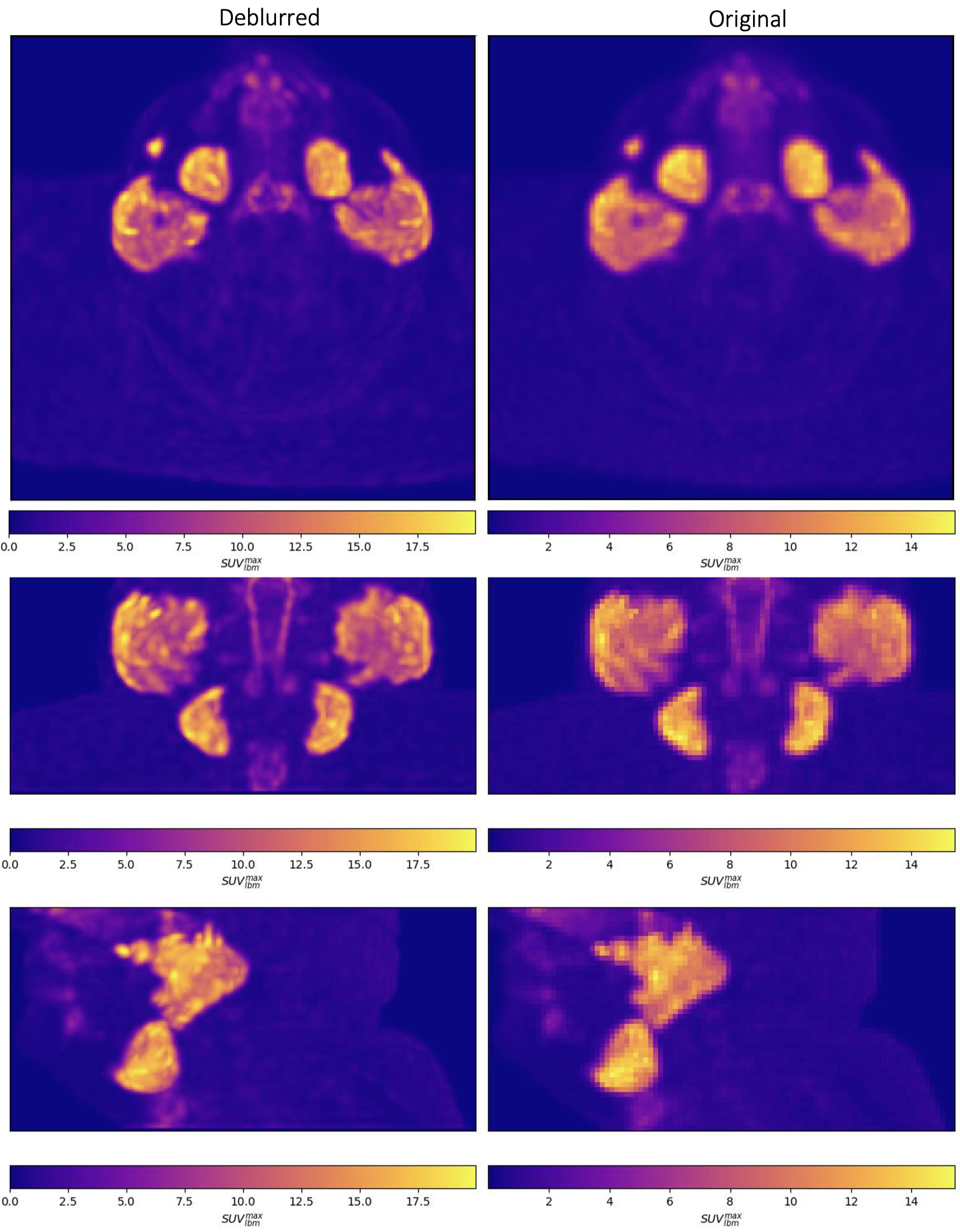}
      \caption[Maximum intensity projections of deblurred and original PSMA PET images]{Maximum intensity projections of deblurred (left) and original (right) PSMA PET images are shown through the head and neck in the axial (top), coronal (middle) and sagittal (bottom) planes.}
      \label{fig:full_body}
    \end{figure}

Maximum intensity projections of $\vect{x}$, $\vect{k}$, $\vect{x} \ast \vect{k}$, and ${y}$ are shown for a single patient in Figure~\ref{fig:x_k_conv_y} and visual comparisons of original and deblurred PET image slices including parotid glands and submandibular glands are shown for 3 different patients in Figure~\ref{fig:par_sm_comp_axial}. Maximum intensity projection images are shown through the head and neck in Figure~\ref{fig:full_body}.
    
\begin{figure}[h]
      \centering
      \includegraphics[width=1\textwidth]{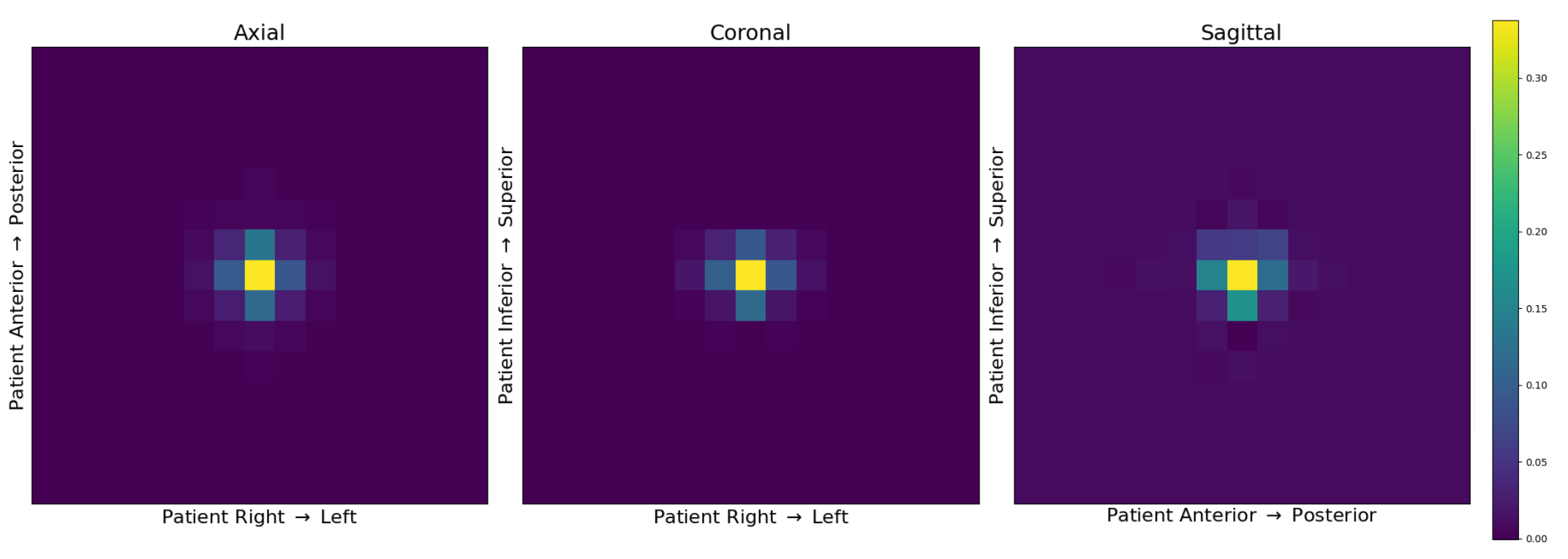}
      \caption{The mean predicted kernel is projected onto the 3 standard patient planes. The predicted blur kernel was found to be mostly symmetric.}
      \label{fig:kernel}
    \end{figure}

    Predicted blur kernels displayed relatively low inter-patient variability, having a mean inner product between different patients of 0.73. Projections of the mean predicted kernel onto the three standard cartesian planes are found in Figure~\ref{fig:kernel}. Predicted kernels were mostly symmetric while having small inter-patient deviations in skewness. The distribution of kernel voxels in various patient directions is summarized in Table~\ref{tab:kernel_dist}, with maximum variation found in the anterior-posterior direction. Kernels were found to display a centred peak, which rapidly fell off to less than 0.01 when two voxels from the center in any direction.

    The mean absolute difference of the total activity found within all voxels of original and deblurred images varied by less than $0.05 \pm 0.07$\%. The predicted blurred image, $\vect{x} \ast \vect{k}$, and the original image, $\vect{y}$, had a MAE and MSE of $0.013 \pm 0.004$ and $0.003 \pm 0.001$ $SUV_{lbm}$, respectively.

\begin{figure}[h]
      \centering
      \includegraphics[width=1.1\textwidth]{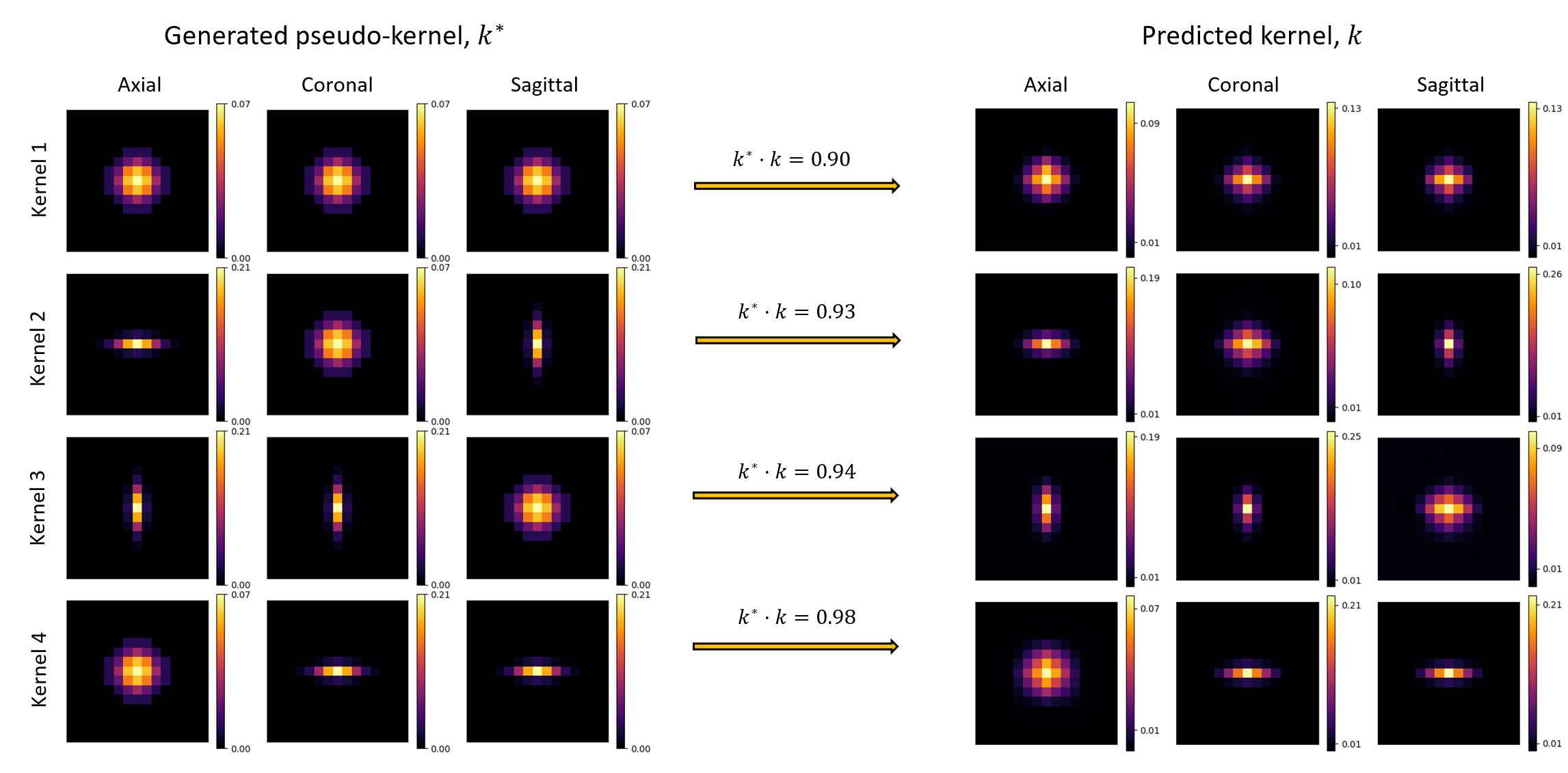}
      \caption{To validate the model's ability to predict accurate blur kernels, 4 variously skewed pseudokernels  were applied to deblurred images before re-running the blind deconvolution. Generated pseudokernels ($k^*$, left) and their corresponding predicted kernel ($k$, right) are shown in separate rows for each of the 4 pseudokernel shapes, in axial, coronal, and sagittal image slices. The dot-product of normalized kernels, $k^* \cdot k$, was used to assess kernel similarity. }
      \label{fig:pseudokernel}
    \end{figure}

    \begin{table}[h]
        \centering
        \captionsetup{justification=raggedright}
        \caption{The fraction of the kernel contained within a plane oriented in each Cartesian direction and intersecting the center of the kernel image, averaged over all patients. For example, the average blur kernel has a sum of 0.52 anterior to the center point.}
        \footnotesize
        \begin{tabular}{@{}ccccccc}
            \br
            Anterior&Posterior&Superior&Inferior&Left&Right\\
            \mr
            $0.52 \pm 0.07$&$0.48 \pm 0.07$&$0.49 \pm 0.05$&$0.51 \pm 0.05$&$0.51 \pm 0.05$&$0.49 \pm 0.05$\\
            
            \br
        \end{tabular}\\
        \label{tab:kernel_dist}

    \end{table}
    Pseudokernels applied to deblurred PET images were predicted well, with mean inner products of predicted and generated pseudokernels above 0.90 for each of the 4 types of pseudokernel applied (regular Gaussian, and skewed Gaussian along 3 patient axes). These results are displayed in Figure~\ref{fig:pseudokernel}
    
\begin{figure}[h]
      \centering
      \includegraphics[width=1\textwidth]{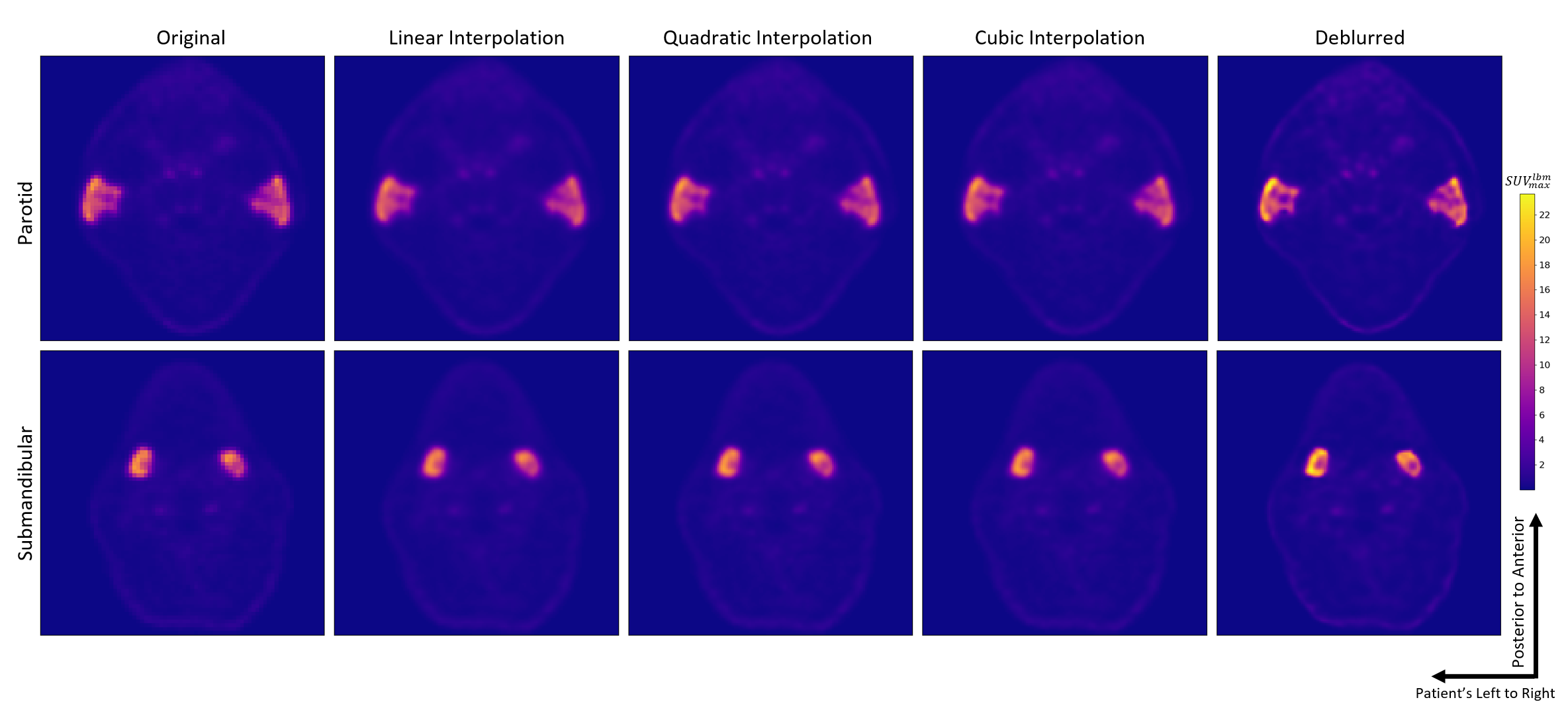}
      \caption{Predicted deblurred images, $\vect{x}$ are twice the original resolution, and we therefore tested whether perceived improvements in image quality were only due to the size difference by upsampling the original image using nearest neighbours, linear, quadratic, and cubic interpolation. The deblurred image (right) appeared to have the best quality.}
      \label{fig:interp_comparison}
    \end{figure}

    Deblurred images appeared to be visually sharper than original images. Since $\vect{x}$ is double the size of $\vect{y}$, we upsampled $\vect{y}$ using nearest-neighbours, linear, quadratic, and cubic interpolation for comparison's sake. Side by side images are found in Figure~\ref{fig:interp_comparison}.

\begin{figure}[p]
      \centering
      \includegraphics[width=1\textwidth]{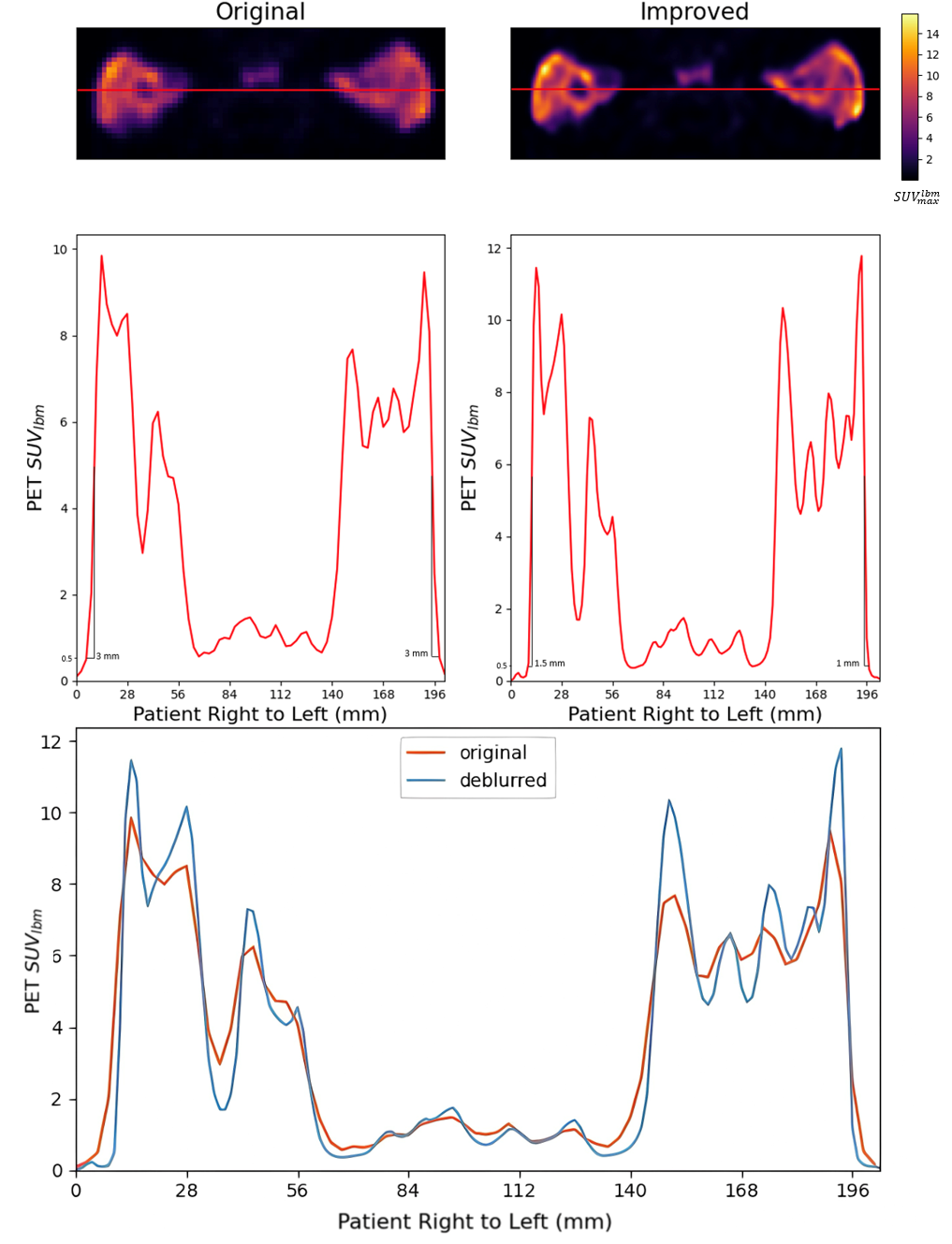}
      \caption{An axial slice through the original (left) and deblurred (right) PSMA PET images, intersecting the parotid glands is shown, along with a red line whose corresponding uptake plot is shown directly below each image. Uptake plots for deblurred and original images are displayed together at the bottom. The deblurred images are found to have smaller partial volume effects, as demonstrated by a more rapid ascent to the full width at half maximum, as shown on lateral edges. In this case, the rise to FWHM was twice as steep in the deblurred images.}
      \label{fig:par_fwhm}
    \end{figure}

\begin{figure}[p]
      \centering
      \includegraphics[width=0.8\textwidth]{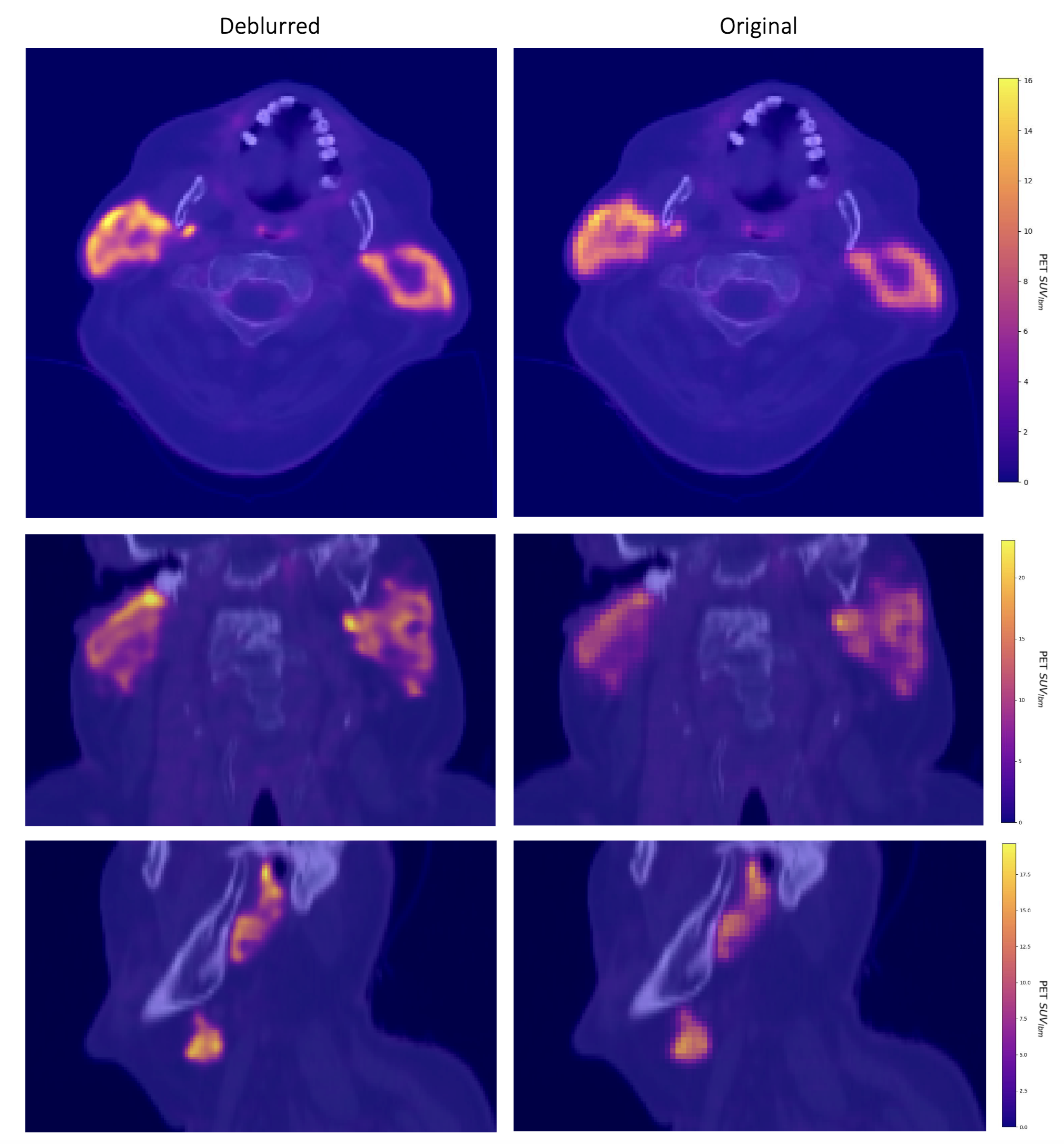}
      \caption{From top to bottom, axial, coronal, and sagittal image slices are shown for both deblurred (left) and original (right) PSMA PET/CT fusion images.}
      \label{fig:fusion}
    \end{figure}

    Cohort uptake statistics within CT-defined salivary glands, calculated using both original and deblurred images, are listed in Table~\ref{tab:gland_stats}. Maximum uptake values in both parotid and submandibular glands were significantly higher in deblurred images ($p < 0.01$), and mean uptake values were insignificantly higher. 
     \begin{table}[h]
        \centering
        \captionsetup{justification=raggedright}
        \caption{PSMA PET uptake statistics in CT-defined salivary glands are summarized using both original and deblurred images.}
        \footnotesize
        \begin{tabular}{@{}ccccccc}
            \br
            &Parotid&& &Submandibular\\
            \\&Original&Deblurred&p-value&Original&Deblurred&p-value\\
            \mr
            $\overline{SUV_{max}}$&$16.2 \pm 4.6$&$21.0\pm 6.2$&p $<$ 0.001&$15.9 \pm 3.9$ & $20.2 \pm 5.2$&p $<$ 0.001 \\
            $\overline{SUV_{mean}}$&$7.1 \pm 1.9$ &$6.7 \pm 1.8$ &p $<$ 0.001& $7.1 \pm 1.8$ & $7.8 \pm 2.2$&p $<$ 0.001 \\

            \br
        \end{tabular}\\
        \label{tab:gland_stats}

    \end{table}
     Uptake plots are shown over corresponding lines through the parotid glands in Figure~\ref{fig:par_fwhm}. The gradient of uptake from 0.5 $SUV_{lbm}$ to the full width at half-maximum (FWHM) is approximately twice as steep for deblurred images than their original counter-parts. The deblurred images display internal heterogeneity within parotid and submandibular glands that appears unresolved in original images. Axial, coronal and sagittal slices of fusion PET/CT images are shown for both deblurred and original images in Figure~\ref{fig:fusion}.

\afterpage{\clearpage}
\clearpage
\subsection{Phantom Analysis}

  \begin{figure}[h]
      \centering
      \includegraphics[width=\textwidth]{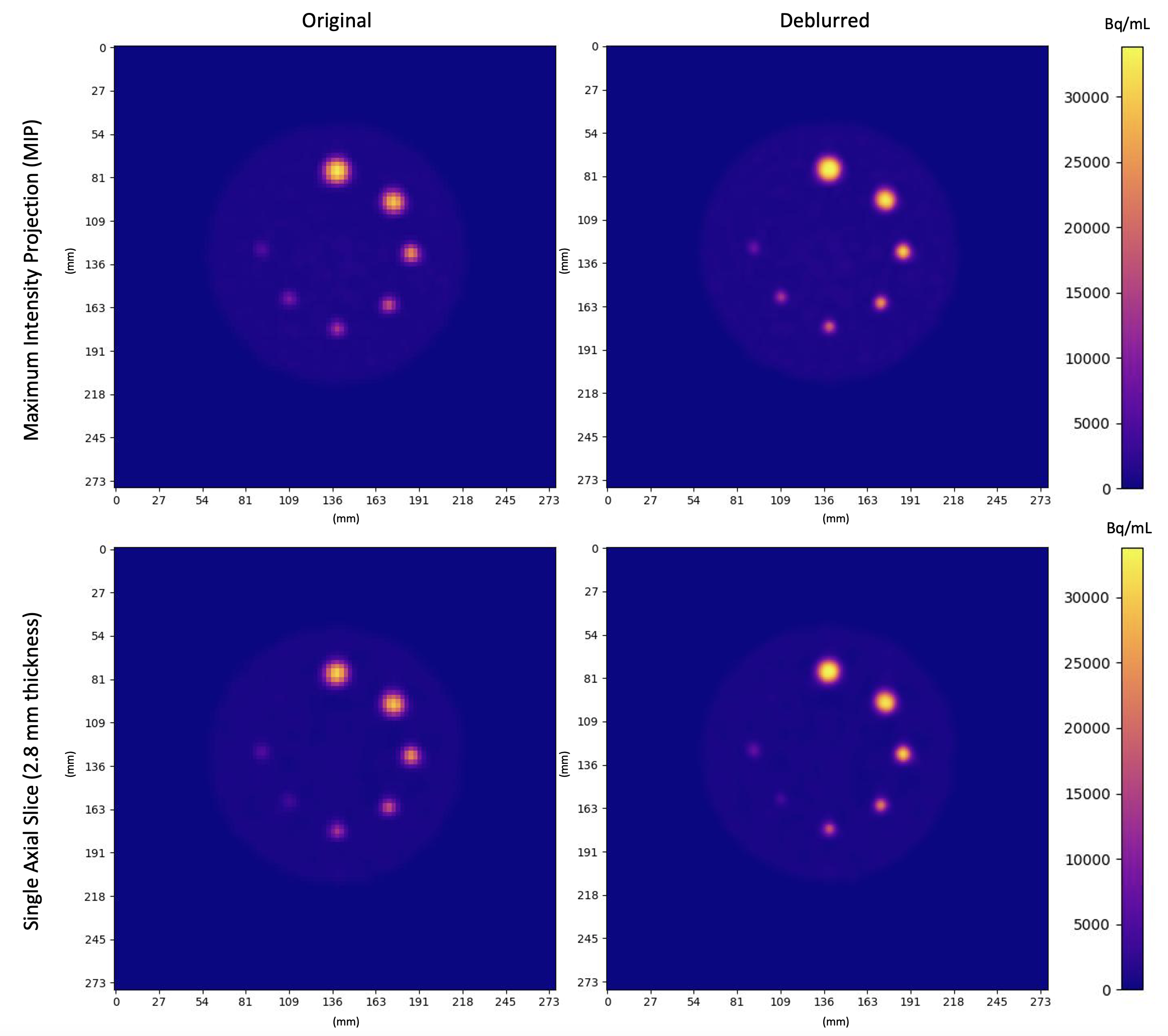}
      \caption{Original and Deblurred PET images of the Q3P phantom. Coplanar spheres containing uniform \textsuperscript{22}NaCl activity concentrations fused in epoxy resin are seen in a single axial slice. Clockwise from the top sphere, diameters were 16 mm, 14 mm, 10 mm, 8 mm, 7 mm, 6 mm, 5 mm. Spheres appear more homogeneous and localized (especially in smaller regions) after deblurring.}
      \label{fig:phantom}
    \end{figure}

\begin{figure}[h]
      \centering
      \includegraphics[width=\textwidth]{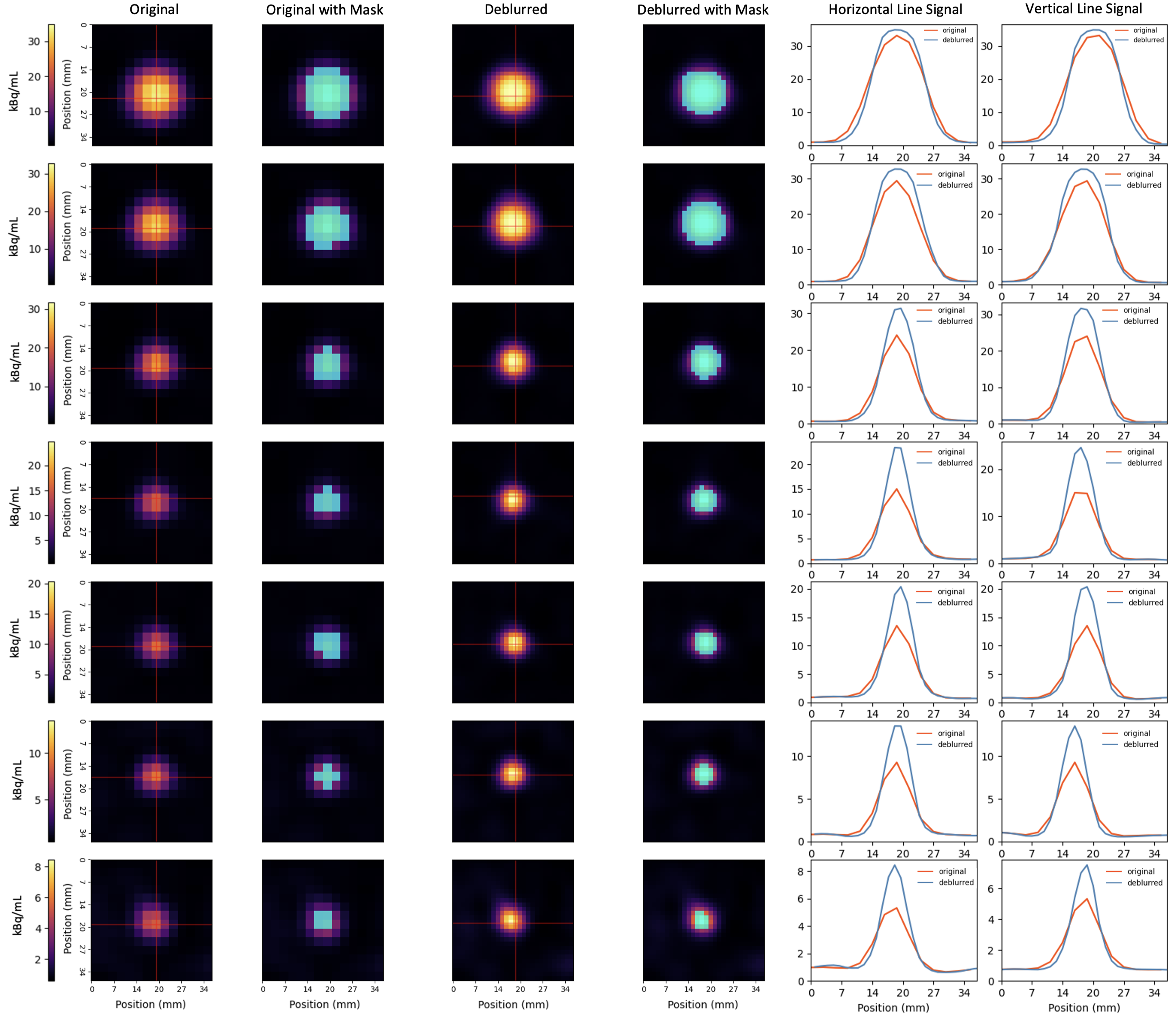}
      \caption{Axial slices of \textsuperscript{22}NaCl infused epoxy spheres in phantom images for original and deblurred images are shown, along with masks corresponding each sphere's best variable threshold for contouring accurate volumes (blue). From top to bottom, sphere diameters are: 16 mm, 14 mm, 10 mm, 8 mm, 7 mm, 6 mm, 5 mm. Activity concentration plots versus distance across each volume in the coronal and sagittal direction are shown to the right, corresponding to red lines shown in PET images.}
      \label{fig:phantom_fwhm}
    \end{figure}

Axial MIPs and slices of original and deblurred phantom images are shown in Figure~\ref{fig:phantom}. Activity concentrations were more localized within spheres after deblurring. This was most pronounced in small spheres ($\leq$ 10 mm). In larger spheres, the activity concentration had higher uniformity after deblurring as seen in signal versus distance plots (across axial slices in both y and x directions) in Figure~\ref{fig:phantom_fwhm}. Figure~\ref{fig:phantom_fwhm} also shows ROI masks defined for each sphere using variable thresholds (Table~\ref{tab:phantom}). Deblurred and supersampling images resulted in ROIs with higher symmetry. The total activity concentration in phantom images was unaltered after deblurring (p $>$ 0.61). 

\begin{landscape}
\begin{table}[h]
        \centering
        \captionsetup{justification=raggedright}
        \caption{Volume and recovery coefficient statistics for \textsuperscript{22}NaCl infused epoxy spheres in original and deblurred images are shown for volumes determined using a single fixed threshold (top) or a variable threshold determined separately for each sphere size (bottom). Deblurring decreased thresholded volume error, and raised recovery coefficients of the mean, median, and maximum activity concentration. Deblurring increased overestimation of the maximum activity concentration in 16 and 14 mm spheres.}
        \footnotesize
        \begin{tabular}{@{}ccccccccccc}
            \br
            \multicolumn{11}{c}{Single Fixed Threshold}\\
            \mr
            Diameter (mm)&\multicolumn{2}{c}{Threshold ($\%$)}&\multicolumn{2}{c}{$V/V_0$}&\multicolumn{2}{c}{$RC_{mean}$}&\multicolumn{2}{c}{$RC_{median}$}&\multicolumn{2}{c}{$RC_{max}$}\\
            &Original&Deblurred&Original&Deblurred&Original&Deblurred&Original&Deblurred&Original&Deblurred\\
            16&42&38&$0.83\pm0.01$&$0.94\pm0.01$&$0.71\pm0.01$&$0.76\pm0.01$&$0.69\pm0.01$&$0.75\pm0.02$&$1.05\pm0.01$&$1.10\pm0.02$\\
            14&42&38&$0.80\pm0.02$&$0.86\pm0.02$&$0.65\pm0.01$&$0.72\pm0.01$&$0.62\pm0.01$&$0.70\pm0.01$&$0.99\pm0.01$&$1.11\pm0.02$\\
           10&42&38&$0.88\pm0.05$&$0.85\pm0.05$&$0.47\pm0.02$&$0.58\pm0.02$&$0.43\pm0.02$&$0.55\pm0.02$&$0.75\pm0.02$&$0.94\pm0.02$\\
            8&42&38&$1.3\pm0.08$&$1.10\pm0.05$&$0.34\pm0.01$&$0.45\pm0.02$&$0.33\pm0.01$&$0.41\pm0.02$&$0.50\pm0.02$&$0.76\pm0.03$\\
            7&42&38&$1.7\pm0.05$&$1.4\pm0.20$&$0.24\pm0.01$&$0.32\pm0.02$&$0.22\pm0.04$&$0.30\pm0.02$&$0.40\pm0.01$&$0.55\pm0.04$\\
            6&42&38&$2.3\pm0.14$&$2.0\pm0.42$&$0.17\pm0.01$&$0.23\pm0.02$&$0.16\pm0.01$&$0.21\pm0.01$&$0.28\pm0.01$&$0.39\pm0.03$\\
            5&42&38&$6.6\pm0.52$&$4.0\pm0.45$&$0.10\pm0.01$&$0.14
            \pm0.01$&$0.08\pm0.01$&$0.13\pm0.01$&$0.16\pm0.01$&$0.24\pm0.01$\\
            \mr
            \multicolumn{11}{c}{Variable Thresholds}\\
            \mr
            Diameter (mm)&\multicolumn{2}{c}{Threshold ($\%$)}&\multicolumn{2}{c}{$V/V_0$}&\multicolumn{2}{c}{$RC_{mean}$}&\multicolumn{2}{c}{$RC_{median}$}&\multicolumn{2}{c}{$RC_{max}$}\\
            &Original&Deblurred&Original&Deblurred&Original&Deblurred&Original&Deblurred&Original&Deblurred\\
            16&36&35&$0.98\pm0.02$&$1.00\pm0.02$&$0.66\pm0.01$&$0.73\pm0.02$&$0.63\pm0.01$&$0.73\pm0.02$&$1.05\pm0.01$&$1.10\pm0.02$\\
            14&36&33&$0.98\pm0.02$&$1.00\pm0.02$&$0.60\pm0.01$&$0.67\pm0.01$&$0.56\pm0.01$&$0.64\pm0.01$&$0.99\pm0.01$&$1.11\pm0.02$\\
           10&41&34&$0.98\pm0.07$&$0.99\pm0.05$&$0.45\pm0.02$&$0.54\pm0.01$&$0.42\pm0.02$&$0.51\pm0.02$&$0.75\pm0.02$&$0.94\pm0.02$\\
            8&54&41&$1.00\pm0.03$&$1.98\pm0.05$&$0.37\pm0.01$&$0.46\pm0.02$&$0.35\pm0.01$&$0.44\pm0.02$&$0.50\pm0.02$&$0.76\pm0.03$\\
            7&52&47&$1.02\pm0.11$&$0.98\pm0.20$&$0.28\pm0.01$&$0.38\pm0.02$&$0.27\pm0.01$&$0.35\pm0.02$&$0.40\pm0.01$&$0.55\pm0.04$\\
            6&59&56&$1.00\pm0.10$&$1.00\pm0.23$&$0.22\pm0.01$&$0.28\pm0.02$&$0.21\pm0.01$&$0.27\pm0.01$&$0.28\pm0.01$&$0.39\pm0.03$\\
            5&79&67&$1.21\pm0.13$&$1.02\pm0.13$&$0.15\pm0.01$&$0.20\pm0.01$&$0.15\pm0.01$&$0.19\pm0.01$&$0.16\pm0.01$&$0.24\pm0.01$\\
            \br
        \end{tabular}\\
        \label{tab:phantom}

    \end{table}
    \end{landscape}

\begin{landscape}
    \begin{figure}[h]
      \centering
      \includegraphics[width=1.5\textheight]{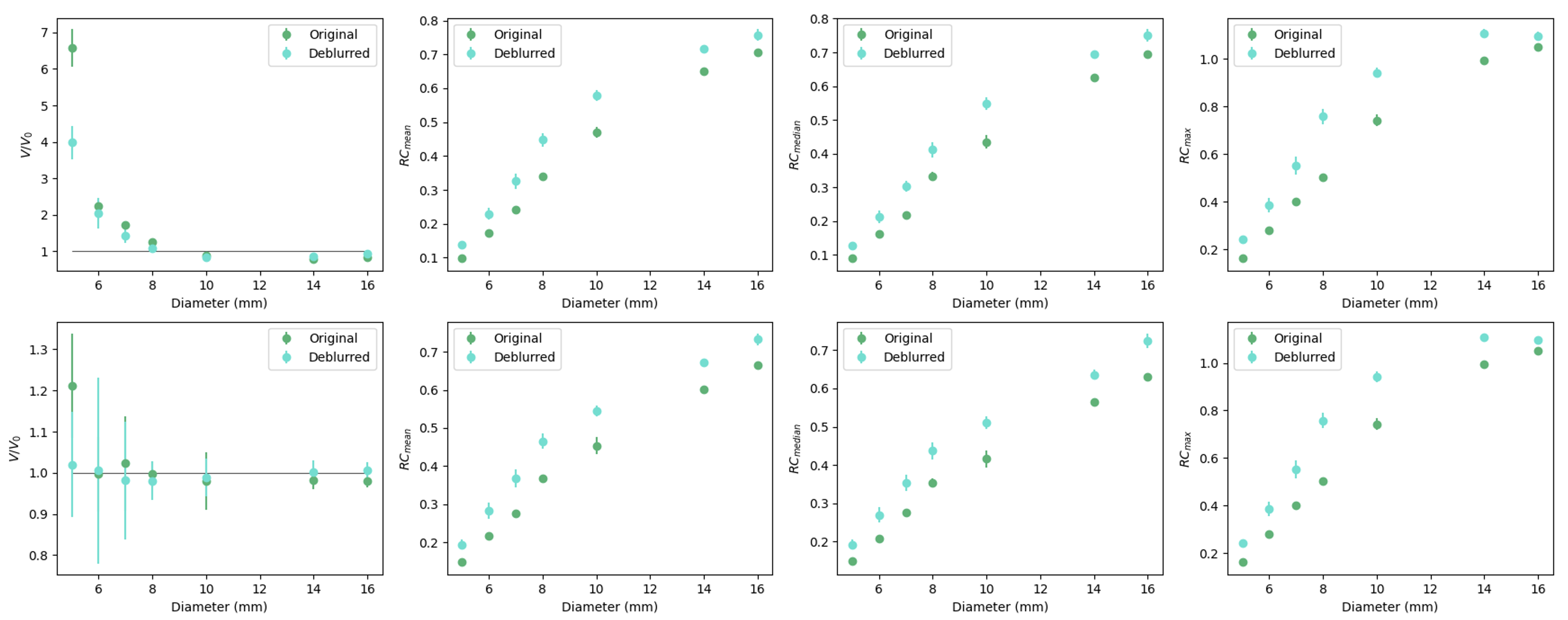}
      \caption{Volume and concentration statistics for \textsuperscript{22}NaCl infused epoxy spheres in phantom images for original and deblurred images. The top row corresponds to sphere statistics determined using volumes defined with a single fixed threshold, while the bottom row corresponds to statistics determined with individual thresholds determined for each sphere size. Spheres defined using deblurred/supersampled images had lower volume discrepancies, and higher recovery coefficients (RCs) of mean, median, and maximum activity concentration. Error bars indicate standard deviations over the 5 consecutively acquired phantom images.}
      \label{fig:phantom_stats}
    \end{figure}
    \end{landscape}

$RC_{mean}$, $RC_{median}$, and $RC_{max}$ increased for each sphere upon deblurring (Table~\ref{tab:phantom}). Small sphere volumes were thresholded with higher volumetric accuracy, and ROI contouring using a single fixed threshold was made more effective for minimizing total volume error over all sphere sizes after deblurring (Figure~\ref{fig:phantom_stats}). In particular, a single fixed threshold characterized small spheres more effectively after deblurring. In both original and deblurred images, the maximum activity concentration in 16 mm and 14 mm exceeded actual activity levels. Deblurring increased this overestimation by approximately 5 \% (Table~\ref{tab:phantom}). Table~\ref{tab:phantom} includes recovery coefficients and volumes of thresholded phantom regions.

Deblurred images of all separate phantom acquisitions were similar, with a mean inner product of $0.93 \pm 0.04$ between distinct kernels. The shape and size of the mean kernel determined for phantom images was similar to that found for patient images. 

\clearpage
\newpage
\section{Discussion}
Based on blind image quality metrics and visual appearance, neural blind deconvolution helped mitigate PVEs and improve overall image quality. Quality improvements were a demonstrated result of the blind deconvolution process, and not only standard upsampling, as shown by comparing results with original images upsampled using various polynomial interpolation orders. The intrinsically low spatial resolution of PET imaging, and consequent PVEs \cite{bettinardi}, render PVE correction a necessary task when quantifying uptake in small image regions. Furthermore, PVE correction of PET images has been shown to increase the statistical significance of correlations between uptake bio-markers and various clinical endpoints and prognostic indices \cite{gallivanone, hatt, hatt_2, hatt_3, ohtaka}.

Predicted blur kernels were consistent with anticipated sizes and shapes, being small (generally no more than 3 voxels across in any direction with values larger than 0.01) and mostly symmetric. By observing uptake fall-off at body borders in original images, it appears that PVEs cause small spill-out to distances larger than one voxel, but neural blind deconvolution cannot detect such small blur components. For model validation, it was essential that predicted kernels shared similarities between patients. The low but non-zero inter-patient variability of predicted kernels was as expected, which can be considered to be an unknown combination of the point spread function, scanner limitations, patient motion and other artifacts.

The model's capacity to predict pseudokernels convolved with previously deblurred images validated its ability to predict various kernel shapes rather than simply Gaussian shapes. All generated pseudokernel shapes were predicted with high accuracy, but it is notable that the highest accuracy (mean inner product = 0.98) was found for pseudokernel's that were stretched onto primarily one axial slice, and all kernel's stretched onto primarily one axis were predicted with a higher accuracy than the symmetric Gaussian kernel. This may be due to the slightly larger slice spacing than pixel spacing (ratio of 1.025), as it is unclear what mechanism in the model architecture would result in increased predictive power of axially flattened kernels. 

The methodology was further validated on phantom images. Activity was more concentrated within spheres and recovery coefficients  were higher after deblurring. Use of a single fixed threshold for defining regions had lower volume error for small regions after deblurring and greater fine detail in small spheres is observable in Figure~\ref{fig:phantom} and Figure~\ref{fig:phantom_fwhm}. These findings, along with improvements in blind image quality metrics and sharpened ROI uptake boundaries support the ability of this method to improve image quality. It is particularly important to further explore the affect of deblurring on the $SUV_{max}$ in lesions, as deblurring led to overestimation of the maximum activity concentration within large spheres. This effect may have been due to the spherical symmetry of phantom lesions, which leads to ambiguity in the determination of a theoretical deblurred image and blur kernel. Overestimation of activity concentrations in large phantom spheres suggest that maximum uptake values in deblurred parotid and submandibular glands may be overestimated. Further exploration of this technique on phantom data with various shapes and sizes could help to understand limitations of this methodology. 

An advantage of neural blind deconvolution over traditional blind deconvolution methods \cite{Gurit} for mitigating PVEs in PET images is that it does not require prior assumptions of the PSF to be imposed. A review of various blind deconvolution algorithms (excluding neural blind deconvolution) applied to natural images \cite{levin} found that the shift-invariant assumption for the blur kernel is often incorrect, which possibly explains a portion of the variance seen in predicted kernels. The proposed method assumes a uniform blur kernel, such that the image can be modelled as $\vect{y} = \vect{x} \ast \vect{k}$. However, it is known that PET image blur is dependent on radial distance from the isocentre (radial astigmatism) \cite{Moses2011,derenzo1981imaging}, so this methodology could likely be improved by modifying the training algorithm to incorporate a spatially varying kernel. However, due to the static nature of the PET scanner, it is likely that this variance is less pronounced here than in cases of natural image acquisition using hand-held devices. 

Previous approaches to super-resolution of PSMA PET images typically seek to reconstruct images of standard quality using lower than standard doses \cite{deng2022,Hu2019, yoshimura2022medical,kennedy2006}. These methods use supervised learning to train models to predict known``ground truth'' high resolution images from their lower resolution counterparts. Other approaches acquire standard images at multiple points of view and attempt to reconstruct single images of supersampled resolution \cite{chang2009}. In contrast, the present approach seeks to supersample reconstructed images of standard quality to higher than standard resolution, using self-supervised learning. Therefore, ground truth images do not exist in the present study for a direct comparison of real and predicted high resolution images. We therefore relied on assessing improvements in blind image quality metrics, and testing kernel similarity between patients and the model's ability to accurately predict pseudokernels. 

To directly compare the present method with previous supersampling techniques \cite{deng2022,Hu2019, yoshimura2022medical,kennedy2006}, a potential study could acquire both low and high resolution PSMA PET images, and then apply the present method to the low resolution images. The acquired ``high resolution'' images could then be compared with predicted values. This allows for direct quantification of supersampling quality.

The performance of neural blind deconvolution on 2D natural images was previously found to out-perform other deblurring approaches \cite{ren_2020, kotera_2021}. A fundamental difference between applying deblurring approaches to PET images vs natural images is that `still' natural images can typically be taken as ground-truth images, while 'still' PSMA PET images have intrinsic blur. This makes it especially challenging to assess deblurring approaches applied to PET images. It was not possible to directly assess the accuracy of predicted blur kernels, as true blur kernels are unknown, and can only be estimated from the scanner's point spread function, which is a source of uncertainty that burdens traditional blind deconvolution \textit{a posteriori} analyses. 

A great advantage of self-supervised learning over supervised learning deep learning algorithms is that a large training set is unnecessary. This is especially important in the case of PSMA PET deblurring, PSMA PET is relatively expensive \cite{Song2022}, making it difficult to acquire large datasets. A self-supervised approach to super-resolution of PET images using generative adversarial networks \cite{Song2020} has had demonstrated success, despite its use of 2D rather than 3D convolutions.

While in the traditional U-Net architecture as well as previous neural blind deconvolution studies, the channel count increases towards deeper regions of the encoder and decoder \cite{ronneberger2015unet, ren_2020, kotera_2021}, we discovered that larger channel counts towards outside regions of the network, decreasing towards the bottom of the encoder and decoder, performed most accurately. Furthermore, including a large number of skip connection channels (64 each) greatly improved performance. This is likely due to increased high-level feature extraction at the top of the network allowing for ultimately better image reconstruction, along with bottle-necking in deeper regions leading to improved learning of important features \cite{Goodfellow-et-al-2016}. Our modelling power was limited by time and GPU resources, and we were unable to test whether scaling all channel counts further could improve results. The model code is available for download online \cite{sample_git}.

Image quality appeared higher within the high uptake regions of salivary glands as well as other low uptake regions. Before adopting the combination of MAE and MSE used in the loss function, we used only MSE as in previous literature\cite{ren_2020, kotera_2021}, but found that this lent itself poorly to the highly skewed distribution of PSMA PET uptake. Adding in MAE allowed appropriate penalization of discrepancy in lower-uptake regions, leading to higher overall quality. We found that decreasing the number of autoencoder layers in $G_x$ from 5 to 4 or 3 had no perceived affect on the model's predictive power, and opened up GPU memory for more channels to be added to the remaining layers. We experimented with various numbers of hidden layers and channel counts in $G_k$ and found that a single layer with 5 times the number of channels in the flattened kernel image worked best. 

PET imaging suffers from intrinsically low resolution, resulting in apparent partial volume effects. We have demonstrated neural blind deconvolution to be a viable post-reconstruction method for mitigating these effects. As opposed to traditional \textit{maximum a posteriori} methods for estimating deblurred images, neural blind deconvolution allows for simultaneous estimation of $\vect{x}$ and $\vect{k}$ without convergence towards a trivial solution. We have built off of previously demonstrated 2D natural image architectures \cite{ren_2020, kotera_2021} to create a suitable architecture for 3D PET images. Inclusion of the CT texture map for steering optimization during early stages helped guide training towards a centered kernel, as opposed to an off-centered kernel and deblurred image, which when convolved, predict a centered image for comparison with the original image. Use of CT images for guiding blind deconvolution relies on accurate image registration. The GLRLM with a long-run emphasis was used rather than the original CT im age, due to its lower contrast which matches PSMA PET better, and due to our previous findings which demonstrated a strong correlation between PSMA PET uptake and the CT GLRLM \cite{sample_2023_hetero}. Initially, a joint entropy loss function was used rather than SSIM, however, this relies on voxel-binning which causes problems during gradient calculation for back-propagation. 
Regularization of the kernel was necessary for avoiding a trivial solution. However, once a MSE penalty was applied to kernel values greater than 0.7, the model converged to non-trivial solutions for all patients. Maximum kernel values were generally less than 0.4, so our kernel constraint did not arbitrarily constrain the maximum value found in predicted kernels to a set value.

This analysis was performed on a dataset of previously reconstructed images, without access to raw scanner data. VPFXS had been previously chosen for clinical image reconstruction over BSREM/Q.Clear \cite{sah2017clinical,teoh2015phantom} for consistency across scanners (which did not all have access to BSREM/Q.Clear reconstruction algorithms). The voxel size is an important acquisition parameter whose modification is sure to affect model performance. The dependence of model performance on voxel size would make for an interesting future study. Many other acquisition and model parameters can be further fine-tuned for optimal performance. Future studies seeking to enhance PSMA PET with neural blind deconvolution should maintain a consistent voxel size across patients for the sake of comparison.

PSMA PET uptake is heavily biased towards the salivary glands than other regions in the head and neck, as seen in figures. A previous study has demonstrated heterogeneous uptake of PSMA-PET in parotid glands \cite{sample_2023_hetero}, and mitigating partial volume effects using neural blind deconvolution appears to make this effect more pronounced. The tubarial glands \cite{valstar_2021} appeared with greater definition in deblurred images, as two distinct regions of high uptake. Maximum uptake regions were better localized in deblurred images, resulting in increased overall maximum values, as well as slight increases in whole-gland means. These results are too be expected, due to mitigated spill-out and spill-in of voxel values around maximum values, and spill-out near gland edges. The total activity in original and deblurred images remained constant, which was necessary in the context of PET imaging. Originally, we included a separate total activity constraint in the loss function, but found this redundant since our fidelity term accomplished this goal.

Supersampling during neural blind deconvolution is not limited to double scaling, and the demonstrated model architecture can easily be amended for accommodating additional resizing. Furthermore, the methodology could be adapted for kernels to be first convolved with deblurred images before downsampling to compute the fidelity loss. Our methodology was limited by time and computation resources.

Deblurring and supersampling PSMA PET images was motivated by the challenge of quantifying uptake in small spatial regions of images, such as the salivary glands, where partial volume effects become increasingly problematic. Prostate specific membrane antigen (PSMA) positron emission tomography (PET) has high ligand accumulation in the parotid glands \cite{trover, israeli, wolf}, and has been suggested to relate to whole-gland functionality \cite{klein, zhao, mohan}. Heterogeneity of PSMA PET uptake trends in salivary glands can be analysed with greater fine detail brought out by deblurring / supersampling. PSMA PET has the unique potential to quantitatively investigate salivary gland physiology, which could potentially lead to better understanding of their functionality as relevant for radiotherapy treatment planning.

PSMA PET is primarily used for detecting and staging prostate cancer, and deblurring/supersampling with neural blind deconvolution could improve fine detail in small lesions, leading to better localization. This methodology has the potential to improve target localization for treatment planning and reduce the rate of false negative lesion detection. Greater localization ability will lead to better estimation of lesion volumes and better monitoring of treatment outcomes. Use of $SUV_{mean}$ over $SUV_{max}$ will likely be necessary for characterising regions of interest as super-resolution algorithms continue to develop and gain clinical confidence, as the $SUV_{max}$ is more sensitive to partial volume effects in small regions. 

\section{Conclusion}
In this work, we have adapted neural blind deconvolution for simultaneous PVE correction and supersampling of 3D PSMA PET images. PVE correction allows for fine detail recovery within the imaging field, such as uptake patterns within salivary glands. Leveraging the power of deep learning without the need for a large training data set or prior probabilistic assumptions, makes neural blind deconvolution a powerful PVE correction method. Our results demonstrate improvements in quality metrics of deblurred images over other commonly used supersampling techniques. The model code is available online for further studies \cite{sample_git}.
\newpage 
\section*{Acknowledgments}
This work was supported by the Canadian Institutes of Health Research (CIHR) Project Grant PJT-162216. 

\section*{Ethical Statement}
This retrospective study was approved by the BC Cancer Agency Research Ethics Board (H21-00518-A001). Participants had provided written consent in prior studies for their data to be further analyzed and for subsequent results to be published.

\section{Data Availability}
Code used in this study is available at: \url{https://github.com/samplecm/neural_blind_deconv_PSMA}
\section*{References}

\bibliography{bib} 
\bibliographystyle{ieeetr}

\end{document}